\documentclass[preprint,12pt]{elsarticle}

\usepackage{amssymb}

\usepackage{siunitx} 
\usepackage{amsmath} 
\usepackage[version=3]{mhchem} 
\usepackage[final]{changes} 

\journal{Journal of Membrane Science}

\begin{document}

\begin{frontmatter}



    \title{Direct 3D observation and unraveling of electroconvection phenomena during concentration polarization at ion-exchange membranes}


    \author[inst1,inst2]{Felix Stockmeier}
    \author[inst1,inst2]{Michael Schatz}
    \author[inst1,inst2]{Malte Habermann}
    \author[inst1]{John Linkhorst}
    \author[inst3]{Ali Mani}
    \author[inst1,inst2]{Matthias Wessling\corref{cor1}}

    \cortext[cor1]{Corresponding author:\\ Email address: manuscripts.cvt@avt.rwth-aachen.de}

    \affiliation[inst1]{organization={Chemical Process Engineering AVT.CVT, RWTH Aachen University},
        addressline={Forckenbeckstraße 51},
        city={Aachen},
        postcode={52074},
        country={Germany}}

    \affiliation[inst2]{organization={DWI - Leibniz-Institute for Interactive Materials},
        addressline={Forckenbeckstraße 50},
        city={Aachen},
        postcode={52074},
        country={Germany}}

    \affiliation[inst3]{organization={Department of Mechanical Engineering, Stanford University},
        city={Stanford},
        postcode={94305},
        state={California},
        country={USA}}

    \begin{abstract}
        A decade ago, two-dimensional microscopic flow visualization proved the theoretically predicted existence of electroconvection roles as well as their decisive role in destabilizing the concentration polarization layer at ion-selective fluid/membrane interfaces. Electroconvection induces chaotic flow vortices injecting volume having bulk concentration into the ion-depleted diffusion layer at the interface. Experimental quantification of these important flow patterns have so far only been carried out in 2D. Numerical direct simulations suggest 3D features, yet experimental proof is lacking. 3D simulations are also limited in covering extended spacial and temporal scales.

        This study presents a new comprehensive experimental method for the time-resolved recording of the 3D electroconvective velocity field near a cation-exchange membrane. For the first time, the spatio-temporal velocity field can be visualized in 3D at multiples of the overlimiting current density. In contrast to today's simulations, these experiments cover length and time scales typical for actual electrodialytic membrane processes.

        We visualize coherent vortex structures and reveal the changes in the velocity field and its statistics during the transition from vortex rolls to vortex rings with increasing current density. The transition is characterized by changes in the rotational direction, mean square velocity, and temporal energy spectrum with only little influence on the spatial spectrum. These findings indicate a more significant impact of EC's structural change on the mean square velocities and temporal spectra than on the spatial spectra. This knowledge is a prerequisite for engineering ion-selective surfaces that will enable the operation of electrically driven processes beyond the diffusion-limited Nernst regime.
    \end{abstract}




\end{frontmatter}

\section{Introduction}

Ion-selective membranes are frequently used in industrial processes for water purification such as electrodialysis (ED)~\citep{Strathmann.2010, Werber.2016} and flow-electrode capacitive deionization (FCDI)~\citep{Gendel.2014, Rommerskirchen.2015, Tang.2020}, or in systems for energy storage like redox-flow battery systems~\citep{Park.2017, Percin.2020}. In these processes, an electric potential between two electrodes acts as the driving force for ion transport through membranes in contact with an electrolyte. For small currents, the current density increases linearly with the applied potential. However, the transport stagnates at a limiting current density i\textsubscript{lim} when concentration polarization (CP) leads to depletion of the Nernst diffusion layer close to the membrane, see Fig.~\ref{fig:StructureOveriV}~a)~\citep{Nikonenko.2014}. This limitation is visible as a plateau region in a current density over potential plot in Fig.~\ref{fig:StructureOveriV}~b).

When increasing the current density beyond i\textsubscript{lim}, overlimiting currents emerge having their origin in a variety of superimposed phenomena, one of them being electroconvection (EC). Electroconvection can be observed as vortices, which convectively mix the boundary layer supplying the depleted layer with the ion-rich bulk solution, see Fig.~\ref{fig:StructureOveriV}~a). The increased ion concentration at the membrane allows for an overlimiting current beyond the diffusion limit, see Fig.~\ref{fig:StructureOveriV}~b)~\citep{Mani.2020}.

Until now, quantitative studies on the hydrodynamics of EC were limited to computationally demanding 2D and 3D direct numerical simulations (DNSs) which unfortunately only offer limited length and time scales~\cite{Druzgalski.2013, Demekhin.2014, Druzgalski.2016, Pham.2016, Guan.2020}: experimental quantification of velocity profiles could only be done using 2D experiments~\citep{Valenca.2015, Valenca.2017,Warren.2021}. In fact, understanding and controlling the hydrodynamics of EC is expected to facilitate mass transport observed as overlimiting currents. \replaced{Multiple studies report membrane modification methods which result in an increased limiting current density or reduced plateau length. Further development of tailor-made modifications as developed by Roghmans et al. will expand the linear regime and might extinguish the limiting plateau~\cite{Roghmans.2019}. This development will enable higher current densities resulting in lower membrane area required to achieve a desired desalination degree: this translates in a desired investment costs reduction.}{Enabling higher current densities results in lower membrane area required to achieve a desired desalination degree: this results in a desired investment costs reduction.} It is therefore obvious that understanding the details of the overlimiting transport, in particular EC, is a prerequisite for process optimization of an electrical-field driven membrane process. Also, understanding electroconvection's intricate details comprehensively is important to tailor membrane surfaces such that overlimiting can be actually utilized in practice.

\section{Background}

In a milestone paper of Rubinstein \cite{Rubinstein.2000}, the authors predicted that the onset of electroconvection can be triggered at lower voltages if the surface of the membrane would be "wavy". This prediction has spurred research in understanding the phenomena itself, but also into the synthesis of new membrane surface topologies to avoid any potential drop in the plateau region prior to the overlimiting currents. With respect to experimental strategies to initiate early electroconvection, membrane modification methods developed by the Kuban research community have proven to be effective \cite{Korzhova.2016, Nebavskaya.2017, Mareev.2018}. As early as 2007, Balster reported that line undulations on the membrane surface normal to the flow direction, having distances in the range of approximately 50-200~\% of the boundary-layer thickness, lead to an earlier onset of the overlimiting current \cite{Balster2007}. Also 2D micropatterns of nanometer-thick lateral polyelectrolyte patches induce electroconvection, i.e. macrosopic electro-osmotic chaotic fluid instabilities \cite{Wessling.2015}. Also Roghmans et al. described that inkjet printed microgel patterns induce electroconvection  \cite{Roghmans.2019}. Yet the work hypothesizes that the surface charge of the patterns may even determine the direction of vortices rendering them either effective or less effective for destabilizing the diffusion boundary layer and reducing the length of the plateau. Clearly, more experimental visualization methods are highly desired to elucidate and quantify the spatio-temporal fluid velocities during electroconvection at the membrane/fluid interface.

\begin{figure*}[!htbp]
    \centering
    \includegraphics[width=1\textwidth]{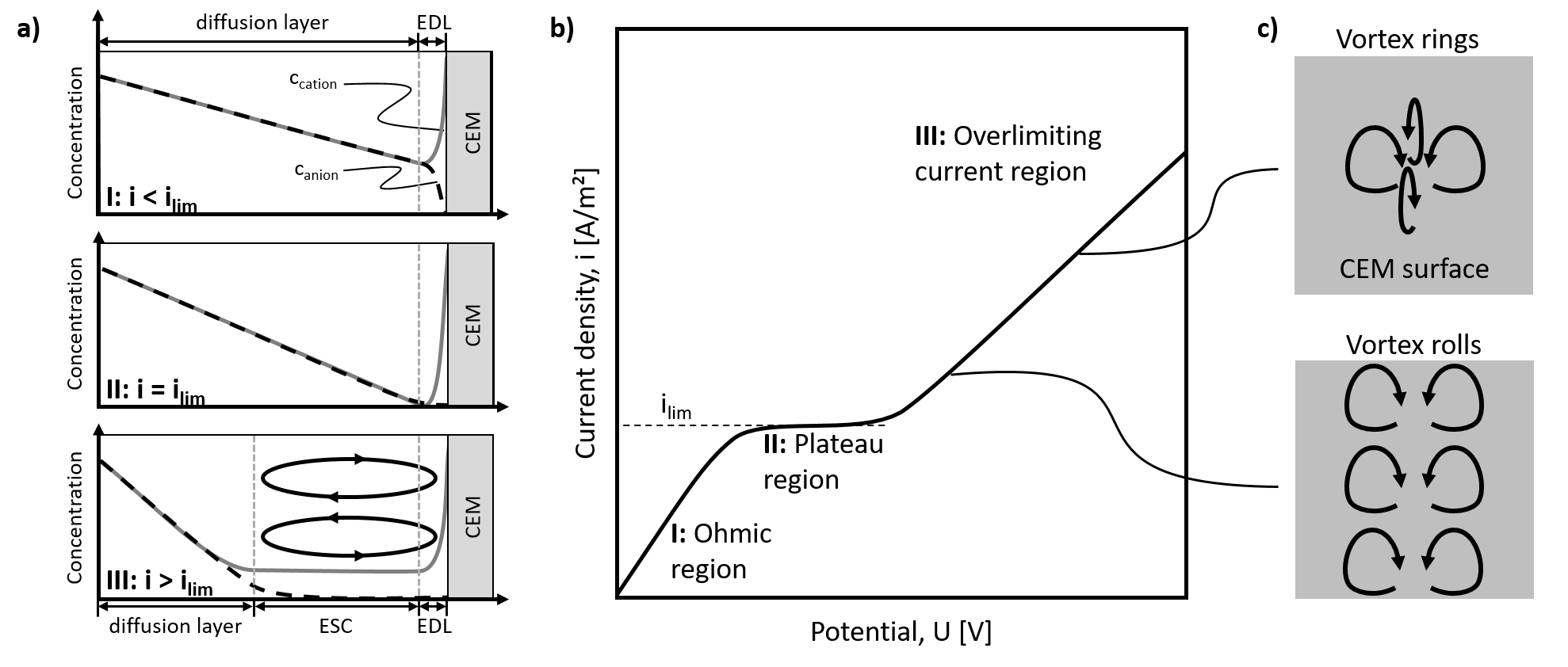}
    \caption{Stages of electroconvection. a) Concentration profiles of anions and cations in front of the membrane for three different regimes (I: $i < i_{\text{lim}}$, II: $i = i_{\text{lim}}$, and III: $i > i_{\text{lim}}$). b) Sketch of the current versus voltage curve for an electrically driven membrane process. The three regimes in a) appear as the Ohmic, plateau, and overlimiting regions. c) Sketches of the different vortex structures developing in the overlimiting regime. First, EC appears as vortex rolls that order in a 2D fashion. With increasing $i/i_{\text{lim}}$, the linear structure of vortex rolls breaks up into 3D vortex rings.}
    \label{fig:StructureOveriV}
\end{figure*}

Historically, our 2D work on proving the existence of electroconvection through visualization with a particle imaging technique \cite{Rubinstein.2008} confirmed the predicted vortex-like features at the membrane/liquid interface with growing vortices at increasing current densities. Subsequently, De Valenca et al. quantified the 2D velocity field of EC in the cross-section of the CP layer ~\citep{Valenca.2015, Valenca.2017}. They correlated the vortex size and fluid flow velocity with the current density through the system. This approach was recently advanced by \citet{Warren.2021} who correlated the effect of the applied potentials \replaced{and electrolyte concentrations}{} to the 2D velocity field of EC. \replaced{They focused their quantification on 2D flow phenomena for the characterization of electro-deposition phenomena by suppressing the evolution of 3D flow with their electrochemical chip's geometry. Therefore,}{ However,} their proposed \replaced{}{electrochemical} chip had an aspect ratio non-existent in common  \replaced{electrochemical devices featuring a long membrane-to-electrode distance compared to the channel's width}{membrane modules}. Other studies also analyzed EC by optically recording tracer substances in 2D but without quantification of velocities~\citep{Rubinstein.2008, Yossifon.2008, Yossifon.2009, Kwak.2013, Kwak.2013b, Kim.2016,Bellon.2019}. Recently, Kang et al. visualized the structural transition of EC in the overlimiting current region with and without flow parallel to a membrane surface with confocal microscopy in 3D~\cite{Kang.2020}. This transition from 2D vortex rolls towards 3D vortex rings, see Fig.~\ref{fig:StructureOveriV}~c), was predicted by Demekhin et al. in the first 3D DNSs of EC~\citep{Demekhin.2014}. Although qualitatively accurate, such confocal microscopy methods do not supply information on the 3D EC velocity field and its energy transfer.

The chaotic 3D velocity field of EC is up to now only accessible via intensive simulations simultaneously solving the Navier-Stokes, Nernst-Planck, and Poisson-Boltzmann equations~\citep{Mani.2020}. In such a simulation, Druzgalski et al. analyzed and compared the statistics of chaotic EC form 2D to 3D results in an attempt to derive energy spectra needed for the development of a statistically averaged reduced-order model~\citep{Druzgalski.2013, Druzgalski.2016}. Such a statistical averaged model would enable low-cost simulations in the overlimiting regime for process optimization~\cite{Mani.2020}. Experimentally, a simple reduced-order parameter approach has been suggested based on fitting the overlimiting currents utilizing a Schmidt number \citep{Stodollick2014}. However, this approach has not been further explored until today. Furthermore, Druzgalski et al. also found qualitative and quantitative differences in their 2D and 3D simulations, emphasizing the need for an new methodology quantifying the 3D hydrodynamics. The required fine resolution in space and time limited their simulations to small geometries and short time spans only. Therefore, the ability to extrapolate their conclusions to the length and time scales of practically relevant devices remains unclear until experimental results of the 3D velocity field are available.\\

To overcome this lack of experimental data \replaced{relevant for industrial electrochemical devices with gap sizes below \SI{1}{\mm}}{}, we developed and report an electrochemical cell allowing the optical recording of the 3D EC velocity field close to a cation-exchange membrane using micro particle tracking velocimetry (\textmu PTV). With this setup, one can now make the important step from 2D to 3D velocity field quantification: at practically relevant length and time scales over a range of overlimiting current densities. We use the velocity data to analyze the structural change of EC from vortex rolls to vortex rings in terms of vortex al direction. Additionally, we evaluate the velocity field statistics and the fluctuating velocity component. \replaced{While our approach excludes the consideration of applied tangential flow, t}{T}his proposed methodology allows to unravel details of EC hydrodynamics experimentally which in future can be compared to simulated velocity statistics when the reported length and time-scales will be accessible through high performance computing efforts.

\section{Experimental Methods}

\subsection{Electrochemical cell}
\label{sec:ECexperiments}

\begin{figure}[htbp]
    \centering
    \includegraphics[width=0.5\textwidth]{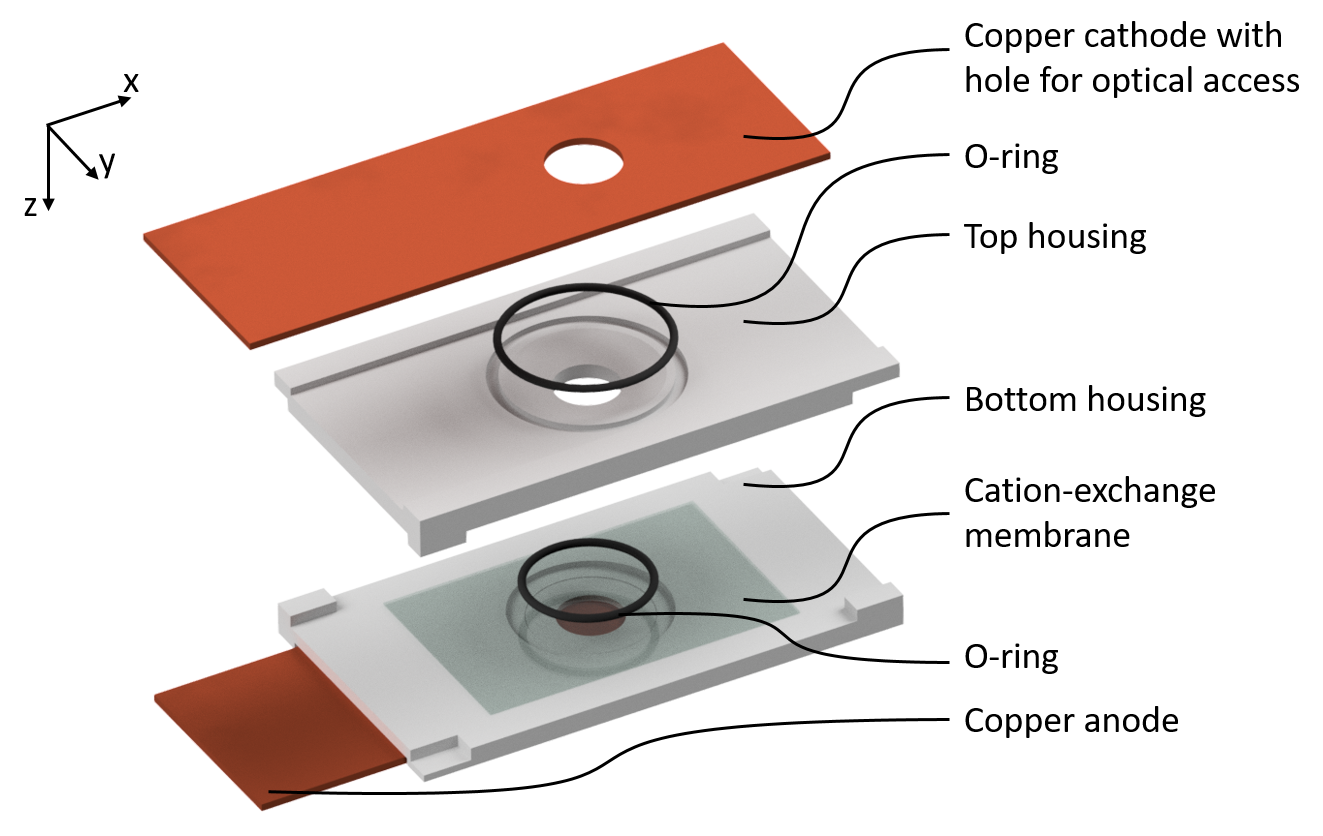}
    \caption{Cell design for electrochemical experiments with simultaneous \textmu PTV recording. The electrochemical cell consists of of two 3D printed housing parts which are sealed with O-rings against two copper electrodes and a Nafion N117 membrane. The cathode has a circular hole with a diameter of \SI{9}{\mm} to allow optical access. The hole is sealed with a microscope glass slide, which is not shown in the rendering, on top of the electrode.}
    \label{fig:cell}
\end{figure}

We designed the experimental cell shown in Fig.~\ref{fig:cell} which is suitable for electrochemical experiments with simultaneous recording of the evolving 3D velocity field using microparticle tracking velocimetry (\textmu PTV). The cell is based on the publication of \citet{Linz.2021} and consists of a Nafion N117 (Chemours, Wilmington, Delaware, USA) cation-exchange membrane sealed between two housing parts. The transparency of the membrane allows for imaging of buoyancy stable EC through the membrane~\citep{Karatay.2016, Valenca.2017b}. Two copper plates (\SI{25}{\mm} $\times$ \SI{75}{\mm} $\times$ \SI{0.5}{\mm}) seal the top and the bottom of the chip and act as electrodes in each half cell. The cathode has a circular hole (d = \SI{9}{\mm}) and is glued to a glass slide to enable optical access. This setup forms electrolyte chambers above and below the membrane, which are filled with \SI{1}{mM} copper sulfate ($\ce{CuSO_4}$) as the electrolyte. The bottom chamber has a height of \SI{0.8}{\mm}, matching the maximum focal depth, and a diameter of \SI{8}{\mm}, resulting in an aspect ratio of 10. Such a large aspect ratio is desired to prevent the confinement of the EC vortices~\citep{Tsai.2004,Davidson.2016}.

A cell comprising two copper electrodes, a cation-exchange membrane, and $\ce{CuSO_4}$ in aqueous solution as electrolyte is a simple and well studied electrochemical system for evoking EC~\citep{Valenca.2015,Gu.2019}. In this system, $\ce{CuSO_4}$ dissociates to $\ce{Cu^{2+}}$ and  $\ce{SO_4^{2-}}$ and both electrodes favor copper deposition and dissolution as faradaic reactions~\citep{Deng.2013}.


During operation, copper is dissolved at the anode, transported through the membrane, and deposited on the cathode. Additionally, this reaction system limits gas evolution at the electrodes and the membrane to a minimum, which ensures stable reaction conditions and clear optical access without gas bubbles.

The copper ions moving towards the cathode result in an ion flux i per membrane area also describes as the current density. The limiting current density of the process, described in Fig.~\ref{fig:StructureOveriV}, can be calculated by~\citep{Valenca.2017}:

\begin{equation}
    i_{\text{lim}} = \frac{c_B}{\delta} \cdot \frac{F \cdot z}{t_M - t_E}
    \label{eq:i_lim}
\end{equation}

According to \citet{Valenca.2017}, the bulk concentration c\textsubscript{B} is approximated with a linear gradient between electrode and membrane as twice the electrolyte starting concentration c\textsubscript{0} in the anode chamber. F is Faraday's constant and z the ionic valence. The transport numbers in the membrane and electrolyte t\textsubscript{M} and t\textsubscript{E} describing the ion velocity difference are 0.96 and 0.4, respectively. For a \SI{1}{mM} $\ce{CuSO_4}$ solution, current densities larger than i\textsubscript{lim} = \SI{0.73}{\A\per\square\m} will result in the EC vortex formation close to the membrane of the anode chamber~\citep{Valenca.2015}.\\

\subsection{Micro particle image velocimetry}

\begin{figure}[htbp]
    \centering
    \includegraphics[width=0.475\textwidth]{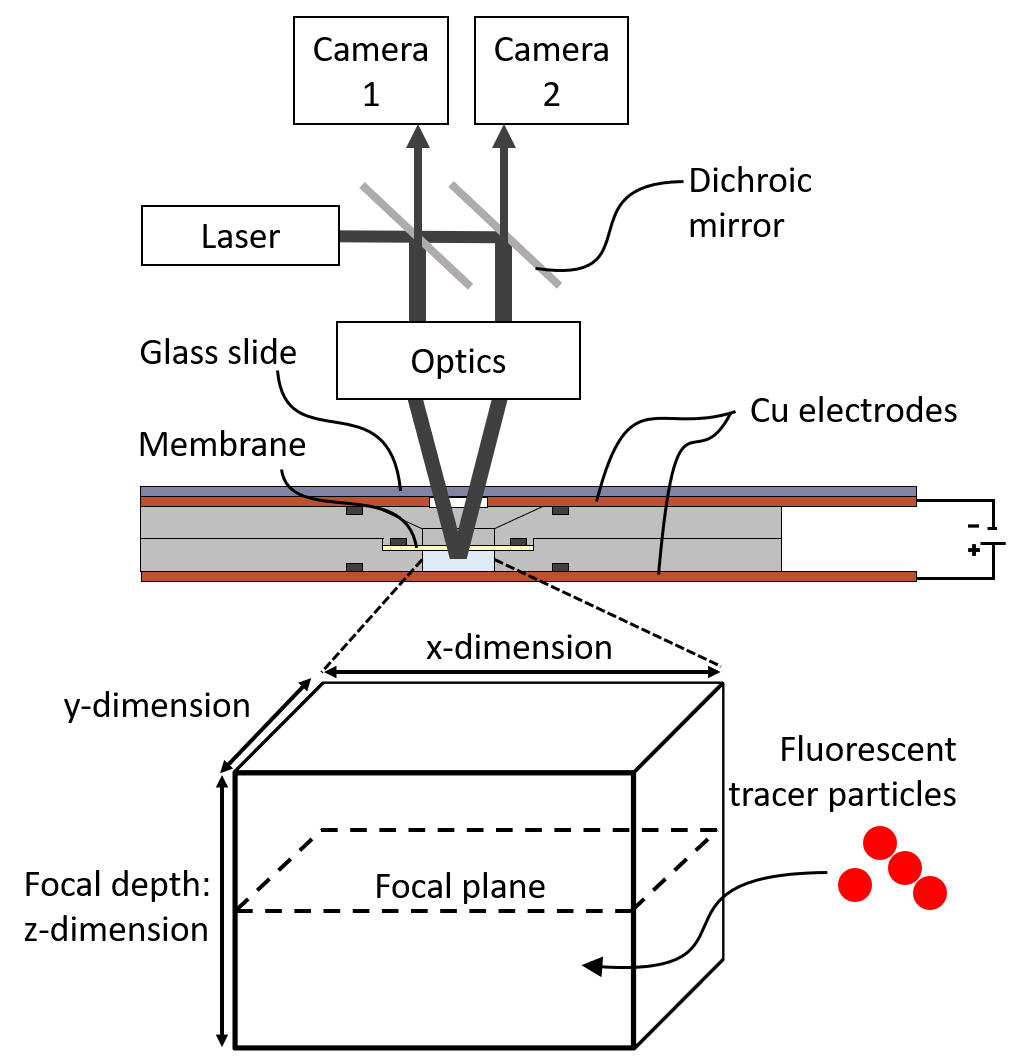}
    \caption{Schematic illustration of the setup for stereo micro particle image velocimetry. The setup for stereo micro particle image velocimetry consists of a laser that illuminates tracer particles in a sample volume through a stereo microscope which splits the laser beam in two separate beams. The fluorescence response returns through the microscope and is recorded by two slightly angled high-speed cameras. Returning laser light is blocked by a high-pass filter. This  technique allows the recording of tracer particles in a 3D volume.}
    \label{fig:PTV}
\end{figure}

We use micro particle tracking velocimetry (\textmu PTV), which is an optical technique for measuring fluid flow in millimeter-scale geometries~\cite{Santiago.1998} to record the velocity field of EC during our experiments. When using this method, tracer particles are illuminated by a high-frequency \SI{532}{\nm} Nd:YAG laser (DM150, Photonics Industries International Inc.) and recorded by two high-speed cameras (Phantom VEO 710L, Vision Research Inc.) through a stereo microscope (SteREO Discovery.V20, Carl Zeiss Microscopy Deutschland GmbH, Germany) containing a 1$\times$ objective (Plan-Aprochromat 1$\times$, Carl Zeiss Microscopy Deutschland GmbH, Germany), see Fig~\ref{fig:PTV}. The cameras record the resulting particle images inside the microscope's focal depth from two different viewing angles. With our setup, the recorded volume at a magnification of $5.12 \times$ and a halfway closed aperture has a size of \SI{4.9}{\mm}$\times$\SI{3.1}{\mm}$\times$\SI{0.8}{\mm} with \SI{1280}{px}$\times$\SI{800}{px} resolution in x- and y-direction. The depth of the recorded volume of \SI{0.8}{\mm} matches the depth of our electrochemical cell's anode chamber. Therefore, it is possible to reconstruct the full velocity field of EC between membrane and anode.

\subsection{Combining electrochemical experiments and velocity recording}

To conduct simultaneous electrochemical and \textmu PTV experiments, the electrolyte in the anode chamber is seeded with 0.001~wt\% inert, fluorescent polystyrene microspheres (Thermo Scientific, Waltham, MA, USA). These tracer particles have a diameter of \SI{3.2}{\micro \m} and a Zeta potential of \SI{-14.9}{\milli \V} measured with a Malvern Zetasizer Nano ZS (Malvern Panalytical Ltd), which are comparable to the study of De Valenca et al.\replaced{ and Warren et al.~\citep{Valenca.2015,Warren.2021}}{~\citep{Valenca.2015}}. \replaced{ Similar to both studies, the electrophoretic velocity of the particles is at least an order of magnitude lower than the velocity gain due to the EC vortices. In contrast to these studies,}{ Different to De Valenca et al.,} the microscope  is focused below the membrane in the middle of the anode chamber enabling to record the particle tracks over the whole z-dimension of that chamber. The experiments are conducted without forced flow and electrolyte movement only results from EC.\\

The chronopotentiometric experiment are conducted at fixed multiples of the limiting current using a potentiostat (Interface 1010E, Gamry, Pennsylvania, USA). For each potential, the time until CP is fully built-up and EC starts to develop can be calculated with Sand's equation~\citep{Krol.1999}:

\begin{equation}
    \tau = \frac{\pi D}{4} \cdot \left( \frac{c_0 z F}{t_M - t_E} \right)^2 \frac{1}{i^2}
    \label{eq:transitionTime}
\end{equation}

Here, c\textsubscript{0} is the electrolyte's initial concentration in the anode chamber. The diffusion coefficient of the electrolyte D = \SI{0.855e-9}{\square \m \per \s} is calculated from both individual diffusion coefficients D\textsubscript{Cu\textsuperscript{2+}} = \SI{0.714e-9}{\square \m \per \s} and D\textsubscript{SO\textsubscript{4}\textsuperscript{2-}} = \SI{1.065e-9}{\square \m \per \s} with the following equation~\citep{Lide.2003}:

\begin{equation}
    D = \frac{(z_+ + z_-) D_+ D_-}{z_+ D_+ + |z_-| D_-}
\end{equation}

The + and - signs indicate the values for cation and anion, respectively. However, EC takes longer than the calculated transition time $\tau$ to reach a steady-state~\citep{Roghmans.2019}. Therefore, we analyze the flow well after the transition time in the steady sections of the chronopotentiometry to ensure evaluation of the steady-state.\\

At 4~$\cdot$~i\textsubscript{lim}, the formation of the EC vortices takes longer than the electrostatic exclusion of tracer particles from the ESC, effectively preventing recording the EC vortex field. In this experiment, the current density is linearly decreased from 8.1~$\cdot$~i\textsubscript{lim} to the desired current density at a rate of \SI{4e-3}{\A\per\square\m\per\s}. Afterward, a steady-state at 4~$\cdot$~i\textsubscript{lim} could be measured. This approach ensures the presence of particles close to the membrane due to previous EC mixing of the domain.\\

When a steady-state in the experiment was reached, the recording of the particle images is started. The flow is recorded for a maximum of \SI{126}{s} at a frequency of \SI{20}{Hz}. The time-resolved recording of the velocity field also enabled the reconstruction of the velocity field during the build-up of EC.\\

\subsection{Velocity processing}

The recorded particle images are analyzed with the tools available in the software DaVis (version: 10.0.5.47779, LaVision GmbH, Göttingen, Germany). In a preprocessing step, static particle signals are removed by subtracting the time-averaged intensity for each pixel. For the reconstruction of the particle tracks, the Shake-the-Box algorithm is used~\citep{Schanz.2013,Schroeder.2015}. However, experiments at current densities below 2~$\cdot$~i\textsubscript{lim} cannot be processed. Here, the particles move too slow and the velocities are too small to be detected.

A fine-scale reconstruction is used for transforming the particle tracks to velocities ($u_x, u_y, u_z$) on a regular grid. The fine-scale reconstruction is performed with the VIC\# method implemented in the DaVis Software~\citep{Jeon.2018,Schneiders.2016}.For our experiments, we chose a grid size of ten voxels, which results in reasonable computing times. Therefore, the final resolution of the reconstructed velocity fields is \SI{128}{px}$\times$\SI{80}{px}$\times$\SI{21}{px} with a voxel size of \SI{38.4}{\micro\m}.

To visualize the vortex structure of EC, DaVis offers plotting coherent vortex structures with the $\lambda_2$-method~\citep{Jeong.1995}. This method computes vortex core lines by evaluating the eigenvalues of the squared and summed symmetric and antisymmetric parts of the gradient velocity tensor. Coherent structures are visualized as isosurfaces at a specific eigenvalue $\lambda_2$.\\

After exporting the velocity field data to MatLab (version: R2019b, The MathWorks Inc.), the mean square velocities is calculated as a spatial and temporal average in the steady-states of the experiments~\citep{Druzgalski.2013}:

\begin{equation}
    \overline{u_i^2} = \frac{1}{N_t \cdot L_x \cdot L_y} \int u_i^2 \cdot dx \cdot dy \cdot dt
\end{equation}

The equation includes the number of time steps N\textsubscript{t} and the mesh points in x- and y-direction L\textsubscript{x} and L\textsubscript{y}.

The fluctuating velocity components are calculated by subtracting the temporal and spatial mean velocity from each velocity data point for evaluation of the temporal and spatial energy spectra. To this data, we apply a Hanning window. The Hanning window or Hann function is a weighting function which is used to reduce the input signal from the middle of the x,y-domain towards its borders~\citep{Essenwanger.1986}:

\begin{equation}
    w_i(n_i) = 0.5 \left( 1 - cos\left( 2 \pi \frac{n_i}{N_i} \right)\right)
\end{equation}

Here, n\textsubscript{i} and N\textsubscript{i} are the array of data points and the number of data points in x- or y-direction, respectively. The 2D window is created with $w_{x,y}(n_x, n_y)~=~w_x(n_x)~\cdot~w_y(n_y)$.\\

\replaced{The fluctuating velocity components are further processed in a spectral analysis enabling the characterization of the energy dissipation in the system.}{}

For  spectral analysis, a Fourier transformation of the fluctuating velocity components over time or space is performed. The velocities, that were 1D Fourier transformed over time $\Tilde{u_i}$, are then converted to temporal spectral energies and averaged in the x,y-planes with~\citep{Druzgalski.2013}:

\begin{equation}
    E(z, \omega) = \frac{1}{2 \cdot N_t \cdot N_x \cdot N_y} \int |\Tilde{u}_x|^2 + |\Tilde{u}_y|^2 + |\Tilde{u}_z|^2 dx dy
\end{equation}
Here, $N_t$, $N_x$, and $N_y$ are the number of data points in time, the x-direction, and the y-direction.

For spatial spectral analysis, a 2D~Fourier transformation of each fluctuating velocity component in each z-plane for each time step is performed. The energy is normalized, and averaged over time:
\begin{equation}
    \begin{split}
        &E(z, k) = \\\
        &\frac{1}{2 \cdot N_t} \cdot \frac{d^6}{(2 \cdot \pi)^3 \cdot L_x \cdot L_y \cdot L_z} \int |\Tilde{u}_x|^2 + |\Tilde{u}_y|^2 + |\Tilde{u}_z|^2 dt
    \end{split}
\end{equation}
The result is integrated over annuli with a radius of $k = \sqrt{k_x^2 + k_y^2}$ and a thickness of one in the Fourier-space (see appendix~\ref{app:int_ref}). Here, d is the mesh spacing, and L\textsubscript{x}, L\textsubscript{y}, and L\textsubscript{z} are the domain sizes in the x-, y-, and z-directions, respectively. This procedure results in energy graphs over the wavenumber~k which is the spatial frequency in x,y-direction. During the integration, the mesh is refined by a factor of ten in both directions to generate smoother graphs.\\

\subsection{Comparability to simulations}

To achive camparability to the 3D DNSs of Druzgalski et al.~\citep{Druzgalski.2016}. We chose process parameters that best match their dimensionless numbers summarized in Tab.~\ref{tab:dimless_numbers}. The relevant dimensionless parameters concerning this comparison include: i) the applied voltage divided by the thermal voltage $\frac{V}{V_T} = \frac{V z e}{k_B T}$, ii) the Schmidt number $Sc = \frac{\mu}{\rho D}$, iii) the electrohydrodynamic coupling constant $\kappa = \frac{\epsilon}{\mu D} \cdot (\frac{k_B T}{z e})^2$, iv) the dimensionless Debye length $\lambda_d = \sqrt{\frac{\epsilon k_B T}{2 c_b (z e L_z)^2}}$, and v) the aspect ratio $a = \frac{d}{L_z}$. Here, $k_B$, $T$, $z$, and $e$ are the Boltzmann's constant, temperature, ionic valence, and the electron charge, respectively. Furthermore, the electrolyte defines the permittivity $\epsilon$, dynamic viscosity \textmu , diffusion coefficient D, and bulk concentration c\textsubscript{b}. Lastly, $d$  and L\textsubscript{z} are the active membrane diameter and the distance between the membrane and electrode, respectively.

Out of these parameters, our main targets were a large channel aspect ratio, a large dimensionless Debye length, and the possibility to conduct experiments at large overlimiting currents, which is proportional to large applied voltages. Advantageous to simulation, our setup allows for dimensions and a Debye length found in practically relevant applications. This difference in Debye length of about three orders of magnitude leads to a smaller extended space charge region in the experiments and, thus, velocity generation closer to the membrane~\citep{Mani.2020}.

\begin{table}[ht]
    \centering
    \caption{Comparison of dimensionless numbers of this study and the reference DNSs~\cite{Druzgalski.2016}. *The applied voltage in our experiments includes the voltage drop at the electrodes, in the membrane, and in the second electrolyte chamber.}
    \begin{tabular}{p{0.25\textwidth} p{0.1\textwidth} p{0.1\textwidth}}
        \hline
        \hline
        Dimensionless number                   & DNS \cite{Druzgalski.2016} & This study                \\
        \hline
        Applied voltage                        & 120 $\frac{V}{V_T}$        & 64 - 362 $\frac{V}{V_T}$* \\
        Schmidt number                         & $1 \times 10^{3}$          & $1.17 \times 10^{3}$      \\
        Electrohydrodyna-mic coupling constant & 0.5                        & 0.14                      \\
        Dimensionless Debye length             & $1 \times 10^{-3}$         & $6 \times 10^{-6}$        \\
        Aspect ratio                           & 6.3                        & 10                        \\
        i/i\textsubscript{lim}                 & 10                         & 1.4 - 10.8                \\
        \hline
        \hline
    \end{tabular}
    \label{tab:dimless_numbers}
\end{table}

\section{Results and Discussion}

Using the electrochemical chip, the movement of tracer particles between the membrane and anode can be followed. Particle tracking velocimetry reveals the 3D velocity field as a consequence of electroconvection at increasing current densities.

\subsection{Evolution of the velocity field with increasing current density}
\label{Sec:iVStructure}

\begin{figure*}[htbp]
    \centering
    \includegraphics[width=0.75\textwidth]{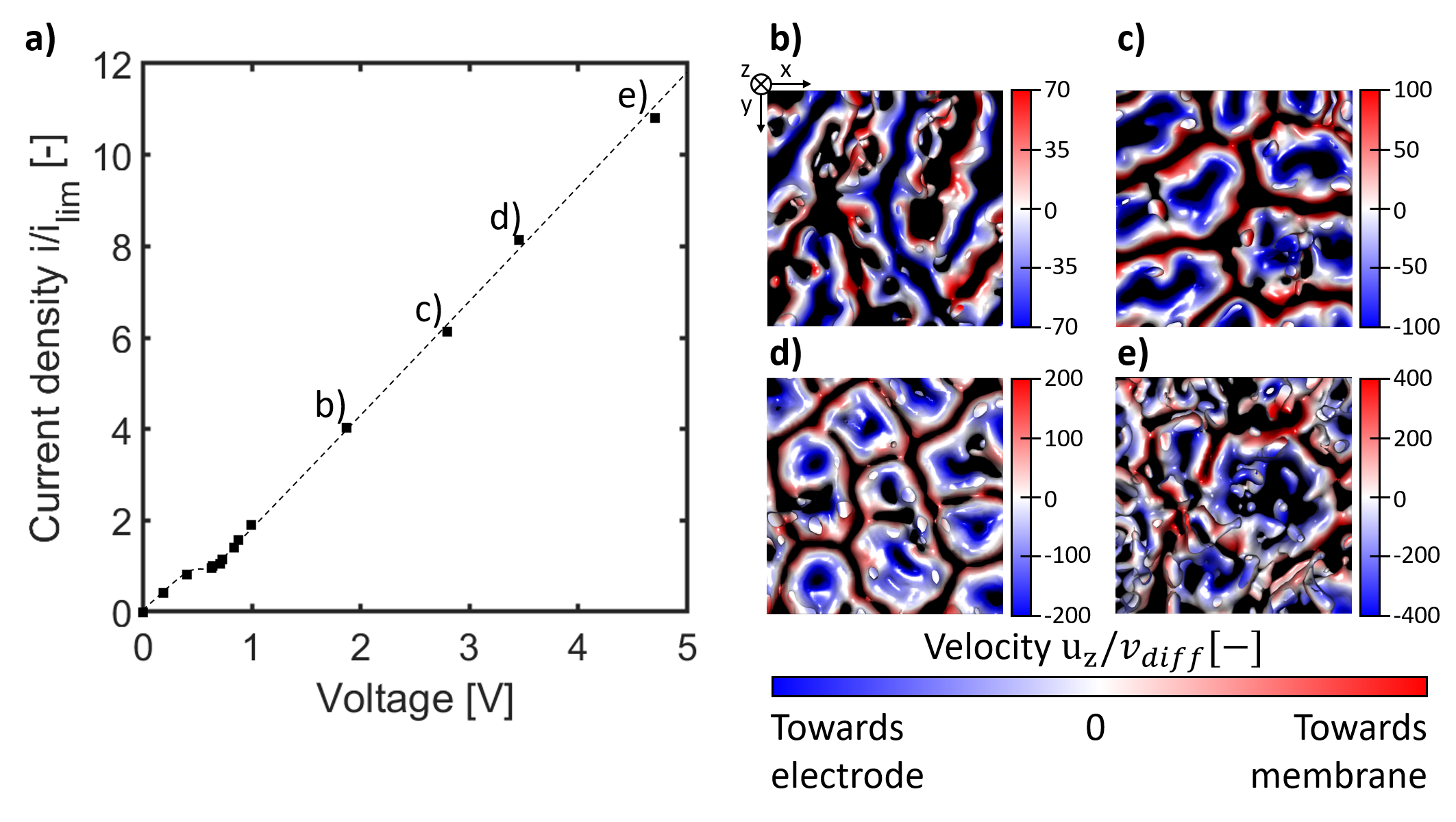}
    \caption{Development of electroconvection in the overlimiting regime. \textbf{a)} The graph shows mean voltages of chronopotentiometry experiments at increasing multiples of the limiting current. The dotted line is drawn for visual guidance. \textbf{b) - e)} Top views on isosurfaces of coherent vortex structures during experiments at \textbf{b)} 4~$\cdot$~i\textsubscript{lim} (at \SI{90}{\s}), \textbf{c)} 6.1~$\cdot$~i\textsubscript{lim} (at \SI{70}{\s}), \textbf{d)} 8.1~$\cdot$~i\textsubscript{lim} (at \SI{70}{\s}), and \textbf{e)} 10.8~$\cdot$~i\textsubscript{lim} (at \SI{81}{\s}) in the steady-state. The velocity is colored in magnitude and direction according to the scale bar from blue to red.}
    \label{fig:2D3D}
\end{figure*}

First, I-V experiments identify the different current regimes in a current density over potential graph at multiples of the theoretical limiting current density ranging from 0~$\cdot$~i\textsubscript{lim} to 10.8~$\cdot$~i\textsubscript{lim}, see Fig.~\ref{fig:2D3D}~\textbf{a)}. The plateau is clearly visible between potentials of about \SI{0.4}{\V} to \SI{0.64}{\V} matching the calculated theoretical limiting current density of \SI{0.73}{\A\per\square\m} (Eq.~\ref{eq:i_lim}). At larger potentials, the system enters the overlimiting current region where EC vortices appear. Graphs of the chronopotentiometric experiments are shown in the appendix~\ref{app:chronos}. The data points shown are time-averaged values of the fluctuating signals having their origin in the dynamic nature of the chaotic flow conditions close to the membrane surface. \\

We record the velocity fields with our setup during experiments at different magnitudes of the overlimiting current. Fig.~\ref{fig:2D3D} b)-e) show snap-shots of the top view on iso-surfaces of coherent vortex structures at 4~$\cdot$~i\textsubscript{lim}, 6.1~$\cdot$~i\textsubscript{lim}, 8.1~$\cdot$~i\textsubscript{lim}, and 10.8~$\cdot$~i\textsubscript{lim}, respectively. With increasing current density, the roll-like structure seen in Fig.~\ref{fig:2D3D}~b) collapses and vortex rings of oval shape emerge, see Fig.~\ref{fig:2D3D}~c). With even higher current densities, the vortex rings first become more regular, see Fig.~\ref{fig:2D3D}~d), before chaotic changing of patterns emerges, see Fig.~\ref{fig:2D3D}~e). While we show the quantification of such 3D velocity field for the first time, this transition from vortex rolls to vortex rings matches the qualitative observations of Demekhin et al.~\citep{Demekhin.2014} and Kang et al.~\citep{Kang.2020}.\\

Additionally to the vortex structure, the measurement of velocity vectors enables the evaluation of rotational directions. At 4~$\cdot$~i\textsubscript{lim}, pairs of counter-rotating vortex rolls are the dominant structure. Fig.~\ref{fig:2D3D}~b) shows that the rolls of a pair connect in the vicinity of other vortex rolls. The velocity between two rolls of a pair is directed towards the membrane. Contrary, the velocity in the centers of vortex rings (Fig.~\ref{fig:2D3D}~c)-e)) is directed away from the membrane. This change in rotational direction indicates that vortex roll pairs do not simply split up in sections and recombine with themselves when transitioning to rings. Two other mechanisms seem more likely: First, vortex roll pairs could split up in sections and recombine but reverse their rotational direction. This change in direction could be induced by the emergence of a smaller vortex ring with a reversed rotational direction in the center of a recombined vortex roll. This smaller ring could increase in size and consume the recombined vortex ring resulting in a ring with a reversed rotational direction. Second, vortex roll pairs could split up at their contact points with other pairs and recombine at the contact point. This observation needs further investigation for a certain explanation.\\

\subsubsection{Particle tracks of vortex rolls}
\label{sec:Rolls}

\begin{figure}[htbp]
    \centering
    \includegraphics[width=0.5\textwidth]{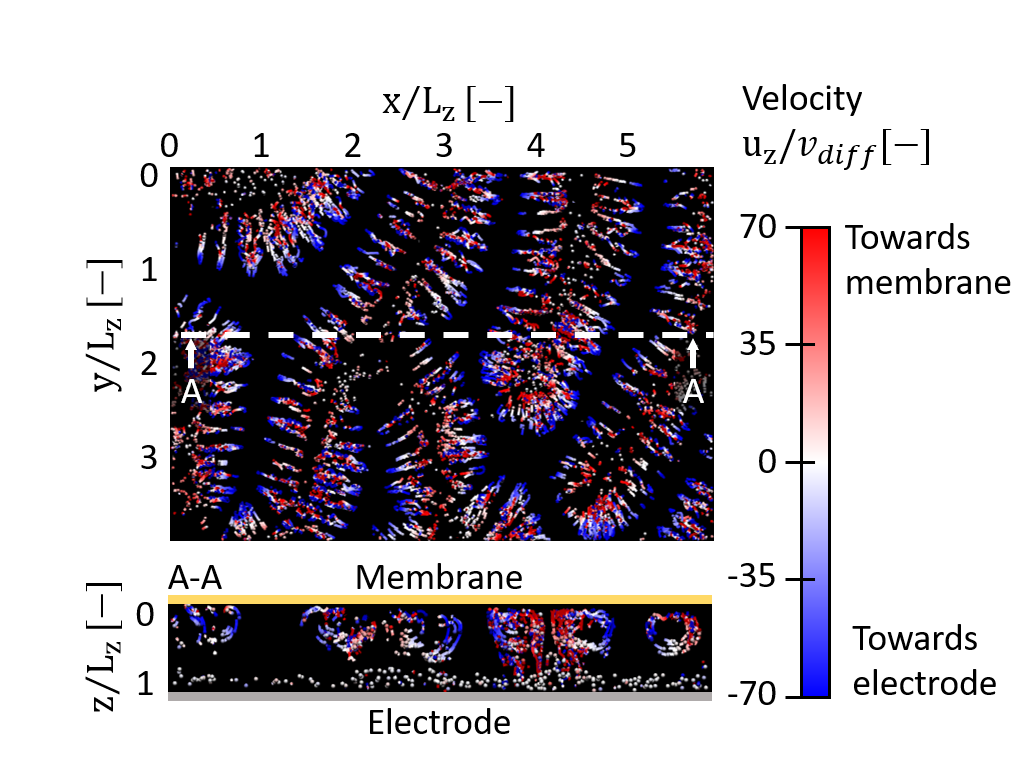}
    \caption{Snapshot of particle tracks for 4~$\cdot$~i\textsubscript{lim} at \SI{90}{\s}. Top view of the particle tracks in the reconstructed 3D volume and a cross-section (A-A) with a track length of 49 time steps (\SI{2.45}{\s}). A-A shows a 2D cross-section of the 3D particle field between the electrode and membrane.  The velocity is colored in magnitude and direction according to the scale bar from blue to red. The axes are scaled by the channel height L\textsubscript{z} = \SI{800}{\micro \m}. Black domains indicate regions where no tracer particles are present due to space charge exclusion as proven by \cite{Davidson.2016}}.
    \label{fig:2DRolls}
\end{figure}

Fig.~\ref{fig:2DRolls} shows the 3D reconstructed particle tracks at 4~$\cdot$~i\textsubscript{lim} which is in the region of vortex rolls. The vortex roll pairs, described in the previous section, are visible in the top view as pairs of particle tracks with opposing rotational directions. De Valenca et al. and Davidson et al. already reported that the slight Zeta potential of the tracer particles leads to their exclusion from locations with high anion concentration~\citep{Valenca.2015, Davidson.2016}. These locations appear as black areas which are void of particles.
Another distinct observation is particles moving on separate orbits along each vortex roll, forming hollow cylindrical structures (Fig.~\ref{fig:2DRolls}). The cross-section A-A shows that the particles are moving on exclusive trajectories. A possible explanation could be the formation of secondary vortices between these separate circles comparable to Taylor flow~\citep{Fane.2009}. These vortices could exclude the particles from the rest of the roll. Another explanation could be the formation of regions of high ion concentration inside the vortex rolls. Such regions would also exclude tracer particles \replaced{similar to the reported phenomenon by De Valenca et al. and Davidson et al.~\citep{Valenca.2015, Davidson.2016}}{}. To identify the validity of these hypotheses, quantitative information on the concentration field would be necessary.

\subsubsection{Veloctity field of chaotic vortex rings}

\begin{figure*}[htbp]
    \centering
    \includegraphics[width=1\textwidth]{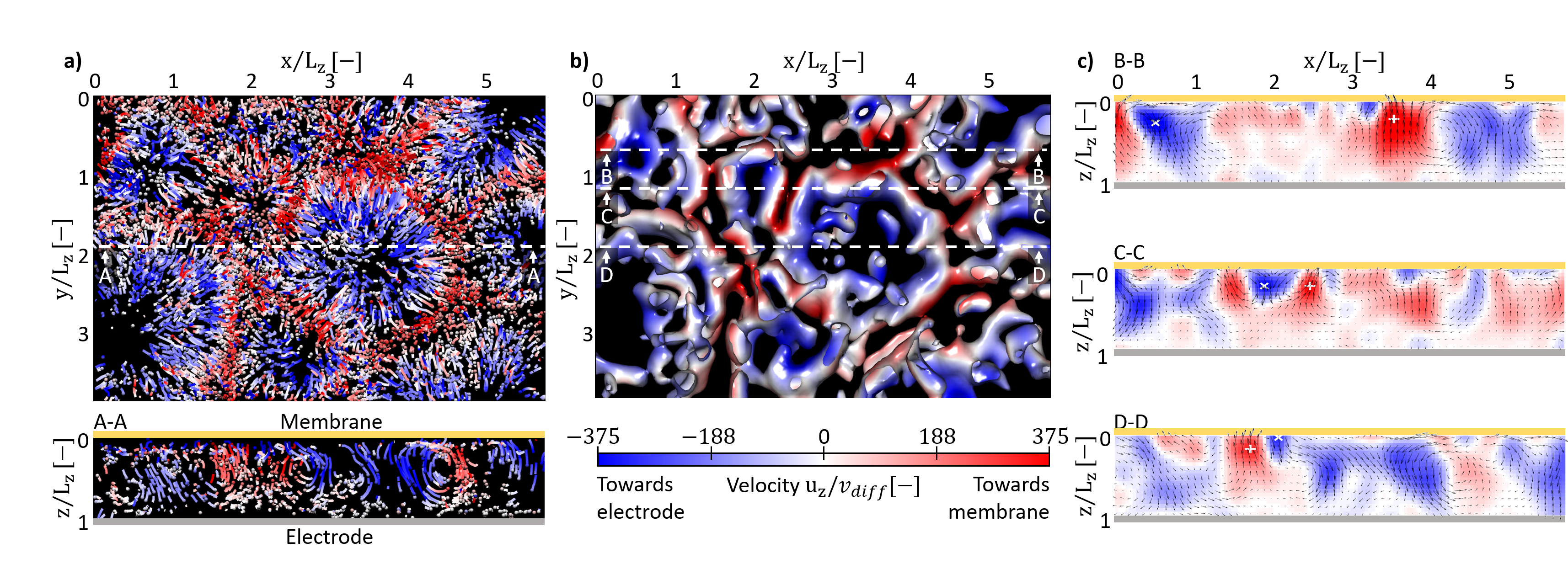}
    \caption{Snapshot of particle tracks and the corresponding velocity correlation for 10.8~$\cdot$~i\textsubscript{lim} at \SI{81}{\s}. a) Top view of the particle tracks in the reconstructed 3D volume with a track length of \SI{1}{\s} . A-A shows a 2D~cross-section of the 3D~particle field between the electrode and membrane. b) Top view of the isosurface of coherent vortex structures in the reconstructed 3D volume showing a chaotic pattern of  vortex rings at \SI{81}{\s}. c) Velocity maps of the cross-section B-B, C-C, and D-D. The white symbols mark the positions of the highest positive (plus) and negative (cross) velocities in each cross-section. The cross-sections A-A and D-D show velocity fields calculated from the particle tracks. Videos of the time-resolved particle tracks and vortex structure can be found as supplementary material.  The velocity is colored in magnitude and direction according to the scale bar from blue to red.}
    \label{fig:particles}
\end{figure*}

Fig.~\ref{fig:particles} shows a snapshot of the steady-state of EC in the regime of chaotic vortex rings at 10.8~$\cdot$~i\textsubscript{lim}. The top view of reconstructed particle tracks in Fig.~\ref{fig:particles}~\textbf{a)}  reveals the appearance of vortex rings with varying diameters. In the cross-section A-A, the particle tracks show circular movement at the borders of vortex rolls, for example, at x/L\textsubscript{z}~=~4.5. These circular tracks strongly resemble structures recorded in the 2D~experiments of Valenca et al.~\citep{Valenca.2015}.\\

For further investigation, the particle tracks are converted to a 3D velocity field. From this velocity field, coherent vortex structures are extracted and displayed as isosurfaces as shown in Fig.~\ref{fig:particles}~\textbf{b)}. The top view of the vortex field structure shows coherent vortex rings as reported by Demekhin et al.~\citep{Demekhin.2014}. The largest displayed velocities seem to appear in the centers of small vortex rings and at contact points of multiple rings. To confirm this observation, cross-sections of the vortex field through such structures are displayed in Fig.~\ref{fig:particles}~\textbf{c)} B-B, C-C, and D-D. The analysis of the cross-sections prove that the largest absolute velocities in the respective planes appear close to the membrane and between coherent vortex rings. These velocities are up to 4.8 times larger than the average absolute velocities in their respective plane. This velocity distribution can be explained by continuity. The area with velocities towards the electrode is proportionally larger than the area with velocities towards the membrane (Fig.~\ref{fig:particles}~\textbf{b)}) which results in higher velocities for the latter.\\

These local velocity hot spots might be connected to local current density hot spots at the membrane. At the positions of high velocities towards the membrane, a larger quantity of high-concentrated electrolyte solution is transported from the bulk to the cation-depleted membrane surface. The resulting local jump in cation concentration will, in turn, lead to a local jump in ionic current through the membrane. Thus, the numerically-predicted formation of current density hot spots~\citep{Druzgalski.2016} could be connected to the velocity hot-spot and verified by our experiments.\\

\subsection{Velocity statistics of electroconvection}

The spatio-temporal 3D velocity information allows to quantify the velocity statistics with increasing current density. This enables to compare the experimental statistics for chaotic vortex rolls to the velocity statistics simulated by Druzgalski et al. from 3D DNSs~\cite{Druzgalski.2016}.

\subsubsection{Comparison of velocity statistics to simulation}
\label{sec:ComVel}

\begin{figure*}[htbp]
    \centering
    \includegraphics[width=1\textwidth]{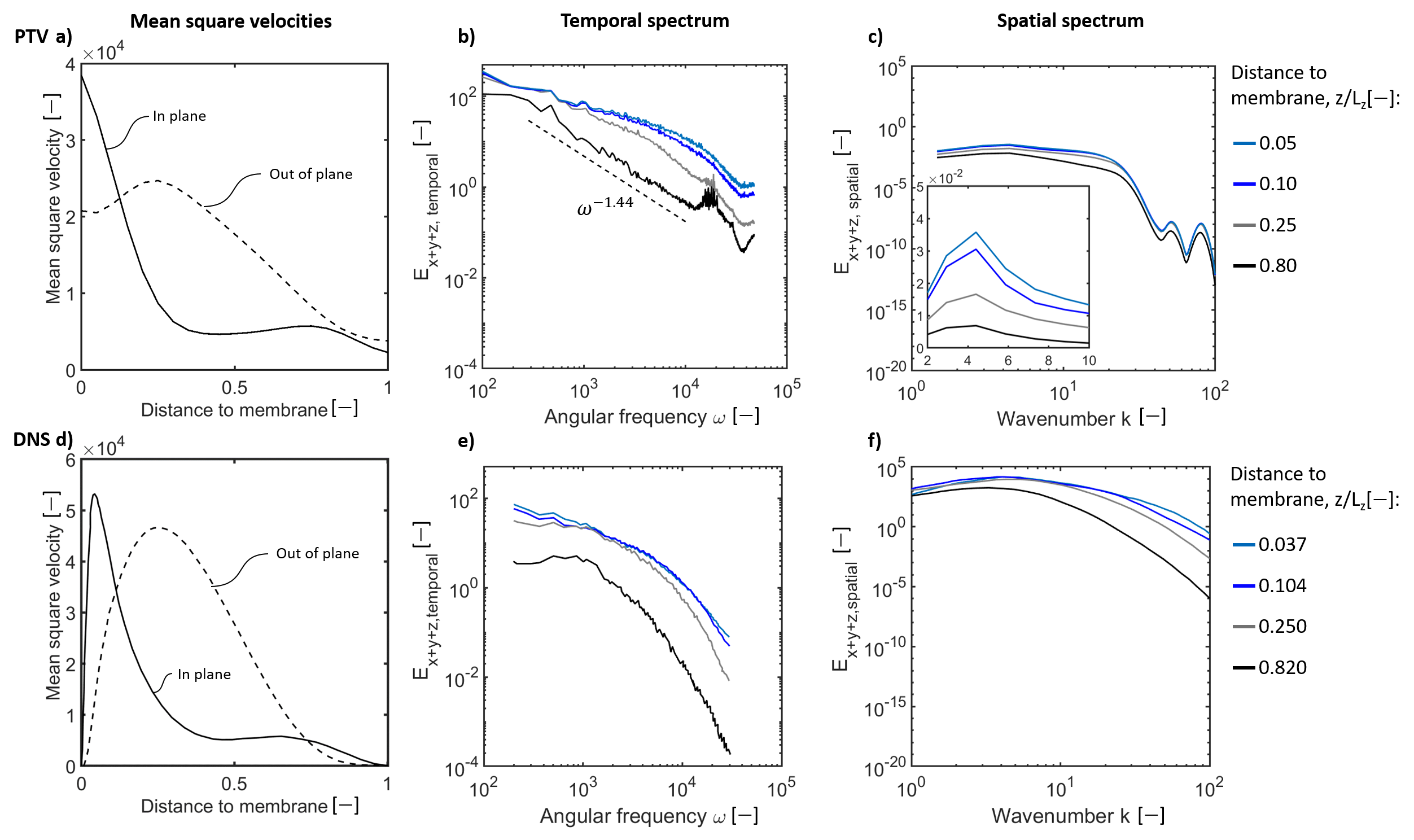}
    \caption{Comparison of mean square velocity, temporal- and spatial energy spectra between 3D~\textmu PTV (upper row) and 3D~DNSs (lower row) ~\cite{Druzgalski.2016}. The graphs present a direct comparison of our 3D~\textmu PTV results at 10.8~$\cdot$~i\textsubscript{lim} (1000 time steps, \SI{74.5}{\s} to \SI{124.5}{\s}) a) - c) to the 3D~DNSs results of Druzgalski et al. d) - f) redrawn from~\cite{Druzgalski.2016} with permission from \textit{Physical Review Letters}. a) and d) show the mean square of one in-plane and the out-of-plane velocity components over the dimensionless distance to the membrane $z/L_z$ with $L_z = 0.8 mm$. b) and e) show temporal energy spectra of the fluctuating velocity components. c) and f) show spatial energy spectra of the fluctuating velocity components. The spectra of 3D~\textmu PTV results b) and c) are plotted for dimensionless distance to the membrane of 0.05, 0.10, 0.25, and 0.80. The linear zoom of the spatial spectrum emphasises the maxima at low wavenumbers. e) and f) are plotted for the normalized distances to the membrane of  0.037, 0.104, 0.250, and 0.820.}
    \label{fig:velocityComparison}
\end{figure*}

Fig.~\ref{fig:velocityComparison} shows the velocity statistics calculated from our experiment with the statistics extracted from the simulations by Druzgalski et al.~\cite{Druzgalski.2016} for 3D DNSs. The statistical analysis quantifies (a) the  mean square velocity and (b) the energy distribution across frequencies.

The mean square velocity for the 3D~\textmu PTV measurements and 3D~simulation~\cite{Druzgalski.2016} are plotted over the normalized distance to the membrane z/L\textsubscript{z} in Fig.~\ref{fig:velocityComparison}~\textbf{a)} and~\textbf{b)}. The mean squares of the in-plane and out-of-plane velocity components show local maxima at about the same distances to the membrane of z/L\textsubscript{z} = 0, 0.25, and 0.75 in the experiment and the simulation. A deviation is identified at the z/L\textsubscript{z} = 0 and 1 positions that can be attributed to the ideal boundary conditions of the simulations compared to the experiments. The maxima positions originate from the vortex generation close to the membrane leading to the first maximum. The second maximum in the out-of-plane component marks half the vortex height with mostly velocity towards or away from the membrane. The last local maximum in the in-plane velocity is located at the average vortex height where the fluid turns its direction. However, the ratio between the amplitudes of both components differs between our measurements and the DNS.

An explanation for the differences between our measurements and the DNSs of Druzgalski et al. can be the discrepancy in the electric double layer thickness by three orders of magnitude, see Tab.~\ref{tab:dimless_numbers}. The consequence of this difference is a decreased ESC region and, thus, the appearance of the maximum in-plane velocities closer to the membrane in the experiments~\citep{Druzgalski.2016}. \replaced{Additionally, t}{T}he velocities in z-direction are averaged in segments of \SI{38.4}{\micro\m} ($\sim$0.05~$\cdot$~z/L\textsubscript{z}) \replaced{height}{length}. This averaging always results in values above zero\replaced{, even}{} in the planes next to the membrane and electrode \replaced{at which a no-slip boundary applies. Furthermore, the use of finite-sized tracer particles excludes volume close to the surfaces}{}. Compared to DNSs, our experiments also include electrodes, the membrane, and a second electrolyte chamber, which influences the electrochemical cell's resistance and chemistry. This comparison proves that the velocity fields and its statistical mean features originating in the EC can be successfully extracted at practically relevant length and time scales.\\

The \textmu PTV velocity field data is used to calculate the temporal spectra in planes parallel to the membrane\replaced{, which indicate the dissipation of energy in the system,}{} shown in Fig.~\ref{fig:velocityComparison}~\textbf{b)}. The energy graphs show similar trends in energy levels compared to the numerical results of Druzgalski et al.~\citep{Druzgalski.2016} in Fig~\ref{fig:velocityComparison}~\textbf{e)}.

Close to the membrane (z/L\textsubscript{z}~=~0.05 and 0.10), the graphs' slope continuously decreases from small to large frequencies. At z/L\textsubscript{z}~=~0.25, a similar energy level is reached in the beginning but shows a more prominent decrease in slope afterwards.
The temporal spectrum closest to the electrode notably start at similar energy levels. Different to the DNSs reference, the energy follows a linear decrease that can be fitted by a power-law of E\textsubscript{x+y+z, temporal}~$\sim \omega^{-1.44}$ between angular frequencies of $\omega$~=~2.8$\cdot$10\textsuperscript{2} and $\omega$~=~1.3$\cdot$10\textsuperscript{4}. Between frequencies of $\omega$~=~1.3$\cdot$10\textsuperscript{4} and $\omega$ = 2.1$\cdot$10\textsuperscript{4}, fluctuations appear followed by a decrease in energy. All graphs show a change in slope above an angular frequency of $\omega$~=~3.5$\cdot$10\textsuperscript{4}.

The fluctuations and changes in slope most probably originate from small-scale noise in the experiments and the measurement error of PTV at small scales~\cite{Herpin.2008}.

The spectra for planes close to the membrane indicate the dissipation of kinetic energy across all frequencies without power-law dependencies, as also reported by Druzgalski et al. for all x,y-planes (Fig.~\ref{fig:velocityComparison}~\textbf{e)})~\citep{Druzgalski.2016}.

The power-law relation close to the electrode can be explained by the forces that evoke EC. The body force that generates EC vortices is located close to the membrane~\citep{Mani.2020}. Therefore,  vortices will propagate from the membrane towards the electrode. The shape of the spectra also indicates the generation of fluctuations of all time scales at the membrane. The fluctuations with small frequencies, e.g. large vortices, could propagate towards the electrode and dissipate their energy, whereas fluctuations with larger frequencies might not reach the regions close to the electrode.\\

The spatial energy spectra calculated from our experiment's velocity fields are displayed in Fig~\ref{fig:velocityComparison}~\textbf{c)}. All graphs follow the same trend at different energy levels. They show a local maximum at a wavenumber of k~$\sim$~4.40. The slope of the graphs decreases at k = 22 with two local maxima above k = 44. These local maxima at high frequencies in the spatial \textmu PTV energy spectra can also be linked to small-scale noise and the measurement error of \textmu PTV for small scales~\cite{Herpin.2008}. The spatial energy spectra reported by Druzgalski et al.~\citep{Druzgalski.2016} are reproduced in Fig~\ref{fig:velocityComparison}~\textbf{f)}. Comparable to our experiment, all graphs show an increase in energy with a maximum at a wavenumber of k~$\sim$~4~\citep{Druzgalski.2016}.

The peak at a wavenumber k~$\sim$~4.40 in the experiment (Fig.~\ref{fig:velocityComparison}~\textbf{c)}) corresponds to a length of 1.43 times the electrode-to-membrane distance (L\textsubscript{z}) carrying the largest energy. This length of 1.43~$\cdot$~L\textsubscript{z} matches well with literature and the diameter of the vortex rings in Fig.~\ref{fig:particles}~\textbf{b)}.

Due to the time-resolved recording of the velocity field, the evolution of the EC statistics over the whole experimental time are additionally analyzed in the appendix~\ref{app:tempDevEC}.\\

\subsection{Development of velocity statistics with increasing current density}

\begin{figure*}[!htbp]
    \centering
    \includegraphics[width=1\textwidth]{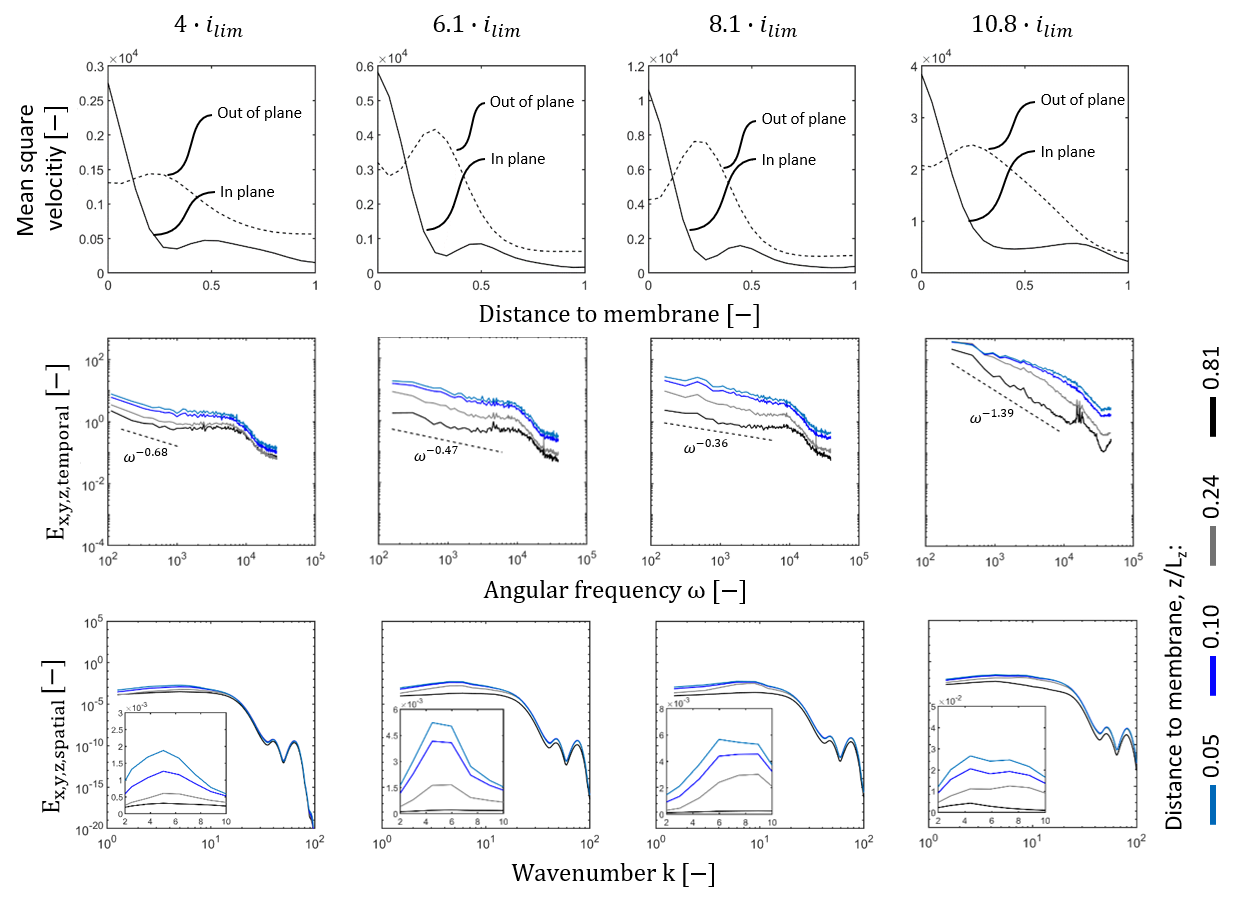}
    \caption{Development of the mean square velocity, temporal and spatial energy spectra in the overlimiting regime for 4~$\cdot$~i\textsubscript{lim}, 6.1~$\cdot$~i\textsubscript{lim}, 8.1~$\cdot$~i\textsubscript{lim}, and 10.8~$\cdot$~i\textsubscript{lim}. The graphs represent an average over 1000 time steps (\SI{74.5}{\s} to \SI{124.5}{\s}). The first row shows the development of the mean square of one in-plane and the out-of-plane velocity components over the plane-to-membrane distance from 3D experiments. The second and third row show the development of the temporal and spatial energy spectra, respectively. Note the changing axes for the graphs showing the mean square velocities and the linear zoom on the maxima of the spatial energy spectra.}
    \label{fig:iVDevECVel}
\end{figure*}

Fig.~\ref{fig:iVDevECVel} shows the mean square velocity, temporal and spatial energy spectra for the experiments at 4~$\cdot$~i\textsubscript{lim}, 6.1~$\cdot$~i\textsubscript{lim}, 8.1~$\cdot$~i\textsubscript{lim}, and 10.8~$\cdot$~i\textsubscript{lim}. All graphs have shapes typical for EC as also seen in Fig.~\ref{fig:velocityComparison}.\\

The overall mean square velocities increase with an increasing current density, as expected due to the higher energy input. The biggest difference appears between 8.1~$\cdot$~i\textsubscript{lim} and 10.8~$\cdot$~i\textsubscript{lim}. Here, the sharp maximum in the out-of-plane velocity and the local maximum in the in-plane velocity seen at 8.1~$\cdot$~i\textsubscript{lim} broaden at 10.8~$\cdot$~i\textsubscript{lim}. Additionally, the local maximum in the in-plane velocity shifts it's position from $\sim$0.5~L\textsubscript{z} to $\sim$0.75~L\textsubscript{z}. This behavior can be explained by the change from the distinct vortex ring pattern into a more chaotic pattern, described in Fig.~\ref{fig:2D3D}, with vortices that span the whole distance between membrane and electrode as seen in Fig.~\ref{fig:particles}.\\

The energy spectra in Fig.~\ref{fig:iVDevECVel} also reflect the overall higher velocities with increased energies for temporal and spatial fluctuations with increasing current densities. Again, the quality of the results at high frequencies in the temporal as well as the spatial energy spectra suffer from small-scale noise and the measurement error of PTV at small scales~\cite{Herpin.2008}.

All temporal energy spectra in Fig.~\ref{fig:iVDevECVel} can be partly fitted by a power-law relation which was also seen in Fig.~\ref{fig:velocityComparison}~\textbf{c)}. The slope of the fitting first increases from $\omega^{-0.68}$ at 4~$\cdot$~i\textsubscript{lim} to $\omega^{-0.36}$ at 8.1~$\cdot$~i\textsubscript{lim} and then decreases to $\omega^{-1.39}$ at 10.8~$\cdot$~i\textsubscript{lim}.

The graphs at 4~$\cdot$~i\textsubscript{lim} additionally show a region of constant energy between $\omega$~=~1~$\times 10\textsuperscript{3}$ and $\omega$~=~1~$\times 10\textsuperscript{4}$ which is less visible at higher current densities. The reason for this steady section could be the absence of information on velocities between the vortex rolls, as seen in Fig.~\ref{fig:2DRolls}. The velocities in these regions potentially fluctuate in small length scales and at high frequencies.

The difference in the temporal energy graphs between 4~$\cdot$~i\textsubscript{lim} and 6.1~$\cdot$~i\textsubscript{lim} coincides with the structural transition from vortex rolls to vortex rings already seen in Fig.~\ref{fig:2D3D}. Moreover, the changes between the graphs at 8.1~$\cdot$~i\textsubscript{lim} and 10.8~$\cdot$~i\textsubscript{lim} match the transition from regular vortex rings to chaotic vortex rings shown in Fig.~\ref{fig:2D3D}.\\

In contrast, the shapes of the spatial spectra do not change significantly between current densities. However, the overall energy increases and the maxima emphasised in the linearly scaled inserts change positions from 5.02~k and 4.48~k to 8.95~k and finally to 4.40~k. These maxima positions reflect the changes in structure that were seen in Fig.~\ref{fig:2D3D}. The correlated diameter of the vortex structures first reduces from 6.1~$\cdot$~i\textsubscript{lim} to 8.1~$\cdot$~i\textsubscript{lim} and then increases again at 10.8~$\cdot$~i\textsubscript{lim}.  The linearly scaled insert at 10.8~$\cdot$~i\textsubscript{lim} shows a broader energy distribution over multiple wavenumbers compared to the graphs at lower current densities. The absence of a sharp maximum indicates that no single dominant length scale exists but a composition of different length scales. This composition is caused by the chaotic change in vortex structure at this large overlimiting current density with the fluctuating size of vortex rings.

Overall, our experiments confirm that large vortices are the dominant actors during EC~\citep{Druzgalski.2016}. The results additionally indicate that an increasing current density seems to have a more considerable impact on the mean square velocities and temporal spectra than on the spatial spectra.\\

\section{Conclusion}

We established the first experimental setup for the time-resolved recording of the 3D velocity field during electroconvection (EC) close to a cation-exchange membrane.
With this setup, Whe EC velocity field was recorded in the steady-state at multiples of the overlimiting current density covering length and time scales found in industrial applications inaccessible by current 3D direct numerical simulations (DNSs).

We visualized coherent vortex structures and revealed the change in vortex structure from vortex rolls to vortex rings along the iV-graph. This change in structure is coherent with the findings reported by Demekhin et al. and Kang et al.~\citep{Demekhin.2014, Kang.2020}. Additionally, our method enables evaluating the rotational direction of vortex structures. We found reversed rotational directions between vortex roll pairs and vortex rings, raising the question of the fluid mechanics during the transition from rolls to rings.

The particle tracks in the regime of vortex rings showed the movement of particles on separate orbits. However, possible explanations for this behavior need further physical evidence.

The velocity measurement in the regime of chaotic vortex rings revealed  local velocity hot spots between the vortex rings directed towards the membrane. These possibly enable the transport of the cation-rich solution at the electrode towards the membrane. This phenomenon could explain the predicted current density hot spots by 3D DNSs~\citep{Druzgalski.2016}.

Further statistical analysis of the velocity field disclosed the nature of energy transfer with a partial power-law behavior in the temporal spectrum. The spatial energy spectrum showed that the length scale carrying the largest energy is 1.43$\cdot$L\textsubscript{z}. These findings showed good agreement of general trends with the reported statistics of Druzgalski et al. for 3D DNSs~\citep{Druzgalski.2016}. Differences were found especially for high frequencies.

Lastly, the evolution of the vortex statistics for increasing limiting current densities was analyzed. The results reveal changes in the mean square velocities and the temporal energy spectra during the transition of EC from vortex rolls over vortex rings to chaotic behavior. The spatial energy spectra showed similar trends for all analyzed current densities. These findings indicated a more significant impact of EC's structural change on the temporal spectra than the spatial spectra.

All in all, the presented results provide a first insight  of the fundamental physics of EC  at practically relevant length and time scales. The experimental technique also builds a foundation to solve the remaining challenges in future detailed 3D studies. Especially the combination of experimental studies, that cover larger scales, and simulations, that more precisely resolve small scales, could lead to significant progress.
Furthermore, the methodology permits studying the impact of membrane surface modifications on the hydrodynamics of EC. These modifications aim to manipulate the 3D~vortex build-up and reduce the limiting current plateau with increasing success~\cite{Balster.2007,Davidson.2016,Valenca.2018b,Benneker.2018b}
Additionally, knowledge of the velocity statistics enables the development of a RANS-like statistically averaged reduced-order model. Such a model will allow for feasibility analysis and optimization of electrically driven membrane processes in the overlimiting regime at a low computational cost. Full comprehension of EC will push the feasibility of desalination processes beyond the limiting current.\\

\appendix

\section{Integration and refinement}
\label{app:int_ref}

\begin{figure*}[htbp]
    \centering
    \includegraphics[width=0.75\textwidth]{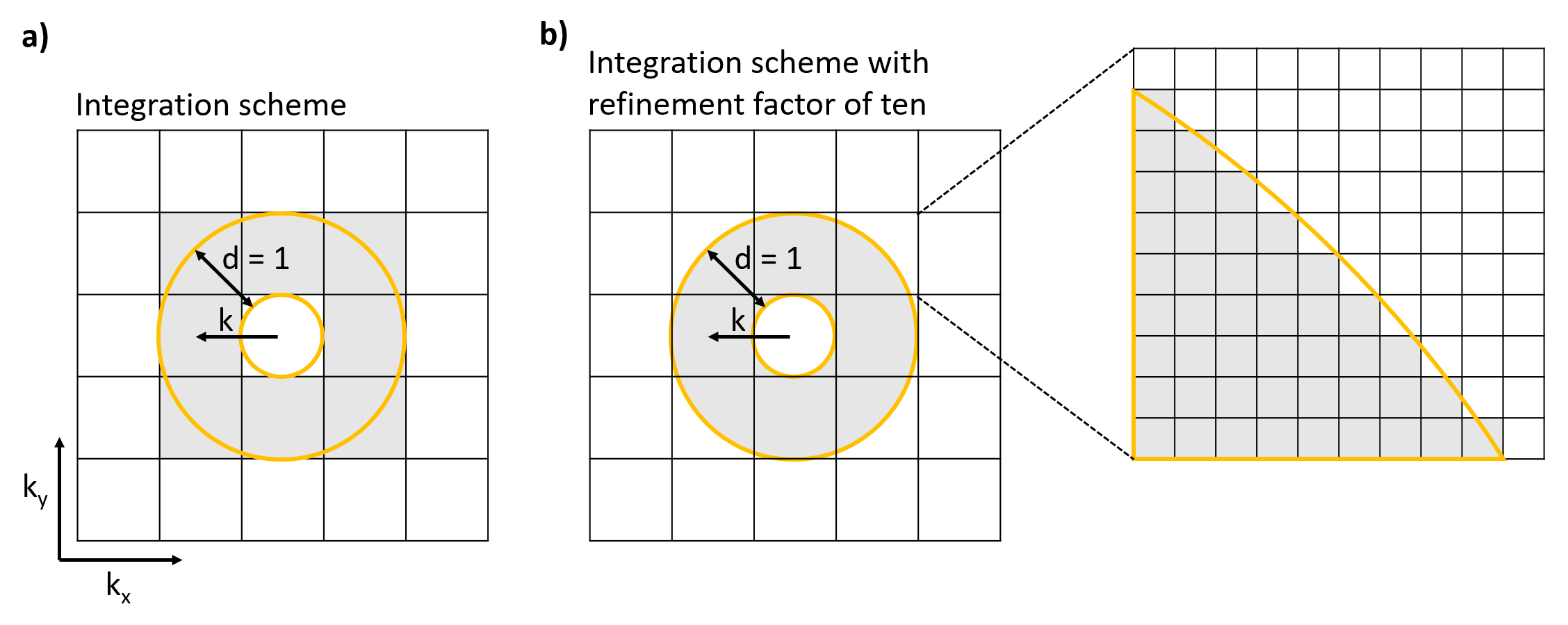}
    \caption{Graphical representation of the spatial spectra integration and refinement scheme. \textbf{a)} Example of the integration for k = 1. \textbf{b)} Example of the refined integration for k = 1.  The detail on the right shows the selected fraction of the data point in one original pixel}
    \label{fig:Integration_and_refinement}
\end{figure*}

Fig.~\ref{fig:Integration_and_refinement} displays a graphical representation of the integration and refinement scheme for the spatial spectra also described by Druzgalski et al.~\citep{Druzgalski.2016}. On the left side (Fig.~\ref{fig:Integration_and_refinement}~\textbf{a)}), the two yellow circles with a difference in radius of one form the annuls for integration. The wavenumber k increases from the center of the k-space in steps of one. The grey highlighted data points show the selection for the non-refined integration. In Fig.~\ref{fig:Integration_and_refinement}~\textbf{b)}, a integration scheme with a refinement factor of ten in both directions is displayed. The refined integration results in a smoother graph especially for small wavenumbers by distributing the information in one cell between two adjacent annuli. The detail on the right shows the selected fraction of the data point.

\section{Chronopotentiometry Graphs}
\label{app:chronos}

\begin{figure}[htbp]
    \centering
    \includegraphics[width=0.5\textwidth]{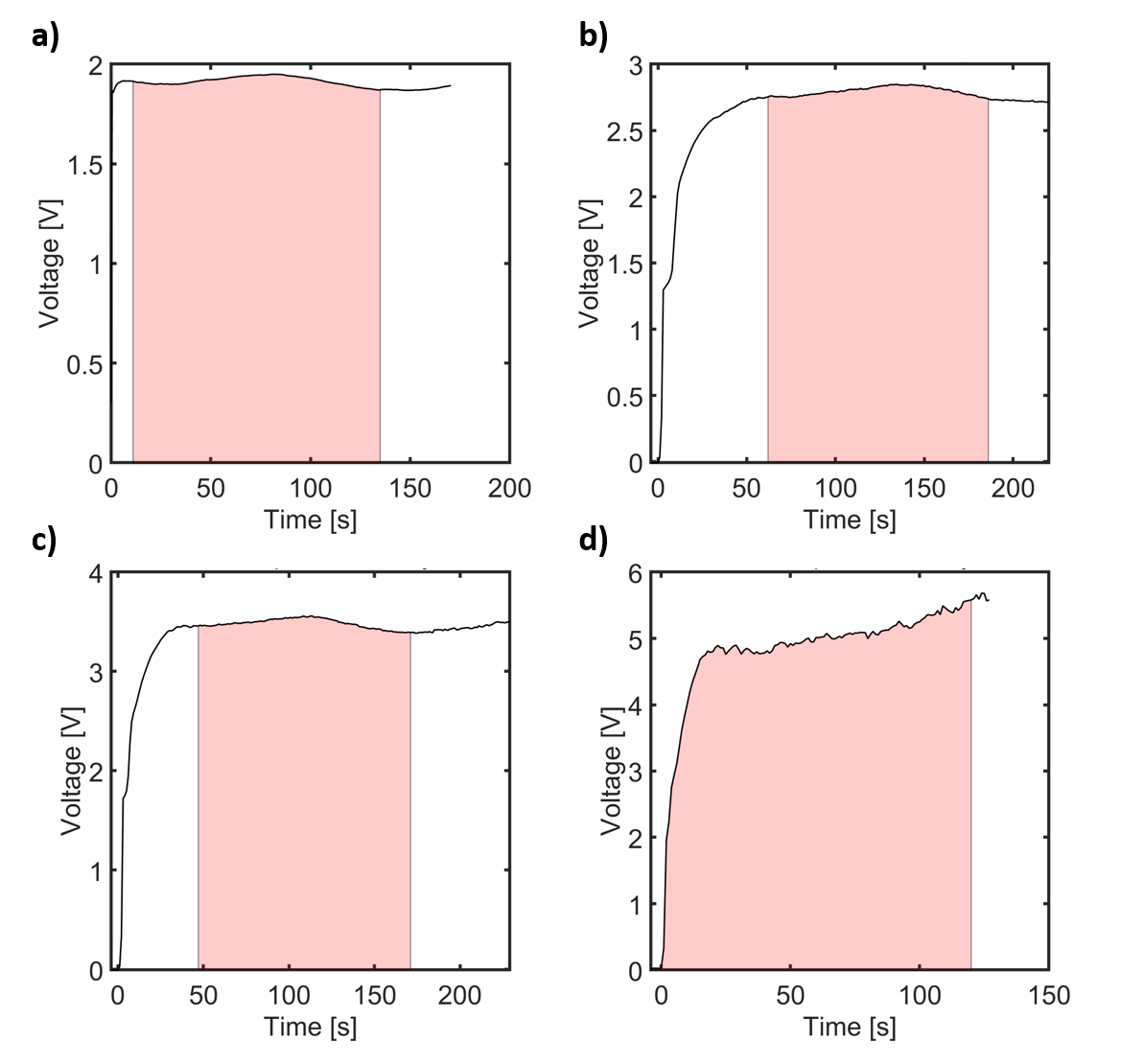}
    \caption{Chronopotentiometry graphs of the experiments at a) 4~$\cdot$~i\textsubscript{lim}, b) 6.1~$\cdot$~i\textsubscript{lim}, c) 8.1~$\cdot$~i\textsubscript{lim}, and d) 10.8~$\cdot$~i\textsubscript{lim}. The red area displays the recorded sections.}
    \label{fig:Chrono}
\end{figure}

Fig.~\ref{fig:Chrono} shows the chronopotentiometry graphs during the experiments at 4~$\cdot$~i\textsubscript{lim}, 6.1~$\cdot$~i\textsubscript{lim}, 8.1~$\cdot$~i\textsubscript{lim}, and 10.8~$\cdot$~i\textsubscript{lim}. The graph in Fig.~\ref{fig:Chrono}~a) directly starts at the steady state potential due to the linearly decreased current density prior to the experiment as described in the main document's method section.

At current densities of 6.1~$\cdot$~i\textsubscript{lim} and 8.1~$\cdot$~i\textsubscript{lim}, the graphs (Fig.~\ref{fig:Chrono}~b) and d)) show a similar trend. After the calculated transition times of \SI{3.94}{\s} and \SI{2.24}{\s}, the systems resistance increases until EC reaches a steady state.

In the experiment at the highest current density of 10.8~$\cdot$~i\textsubscript{lim} (Fig.~\ref{fig:Chrono}~d)), the potential increases over the course of the measurement. This behavior indicates increasing resistance of the system due to chemical changes or gas bubble formation. Both phenomena could originate from possible water dissociation at the membrane and the electrodes at this large potentials. The resulting $\ce{OH^-}$ ions possibly form solid $\ce{CuO}$ or $\ce{Cu(OH)_2}$ scaling at or inside the membrane~\cite{Chang.2010}. In our experiments, we did not observe any visual changes of neither the membrane nor the electrodes. Excess $\ce{H^+}$ and $\ce{OH^-}$ would form $\ce{H_2}$ and $\ce{O_2}$ gas bubbles at the electrodes. However, gas bubble formation was only observed for experimental times larger than \SI{500}{\s}.

\section{Temporal Development of Chaotic Electroconvection}
\label{app:tempDevEC}

\begin{figure}[htbp]
    \centering
    \includegraphics[width=0.5\textwidth]{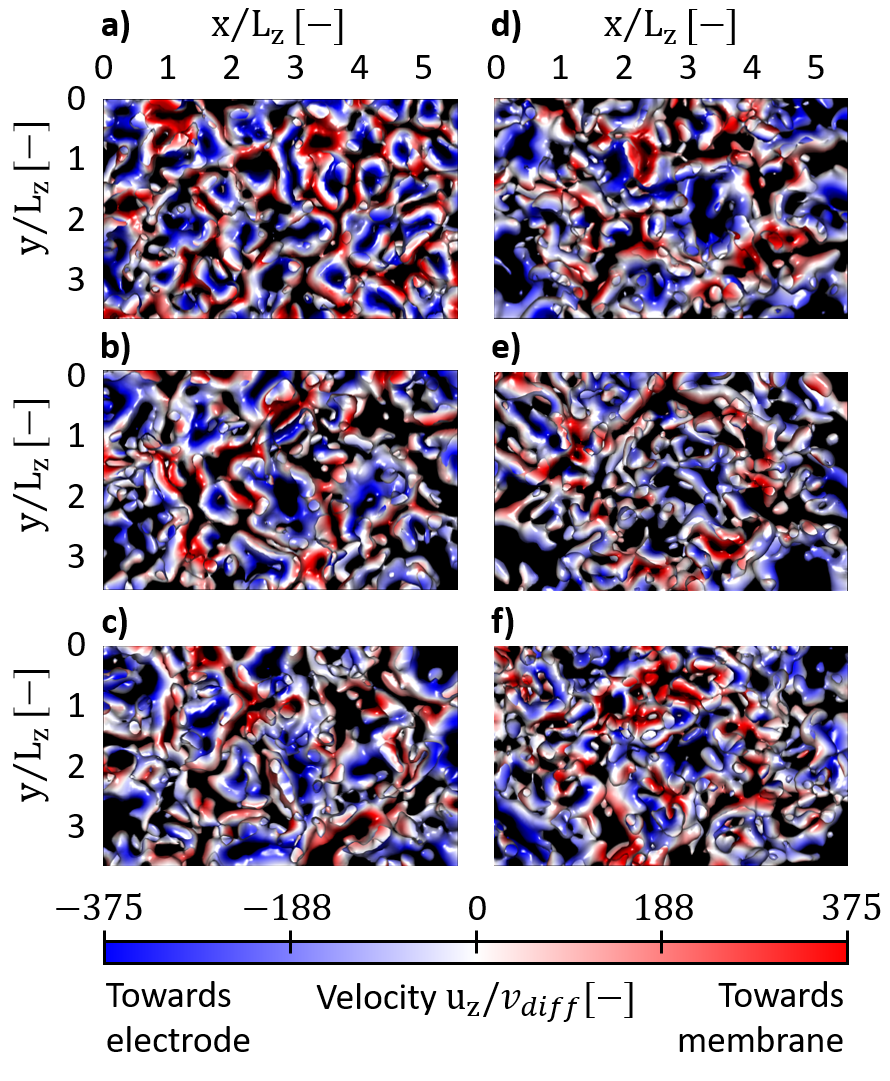}
    \caption{Temporal development of the EC vortex structure for 10.8~$\cdot$~i\textsubscript{lim}. The top views of the structure of coherent vortices is depicted at \textbf{a)} \SI{20}{\s}, \textbf{b)} \SI{40}{\s}, \textbf{c)} \SI{60}{\s}, \textbf{d)} \SI{80}{\s}, \textbf{e)} \SI{100}{\s}, and \textbf{f)} \SI{120}{\s} of the experiment.}
    \label{fig:tempDevEC}
\end{figure}

\begin{figure*}[htbp]
    \centering
    \includegraphics[width=1\textwidth]{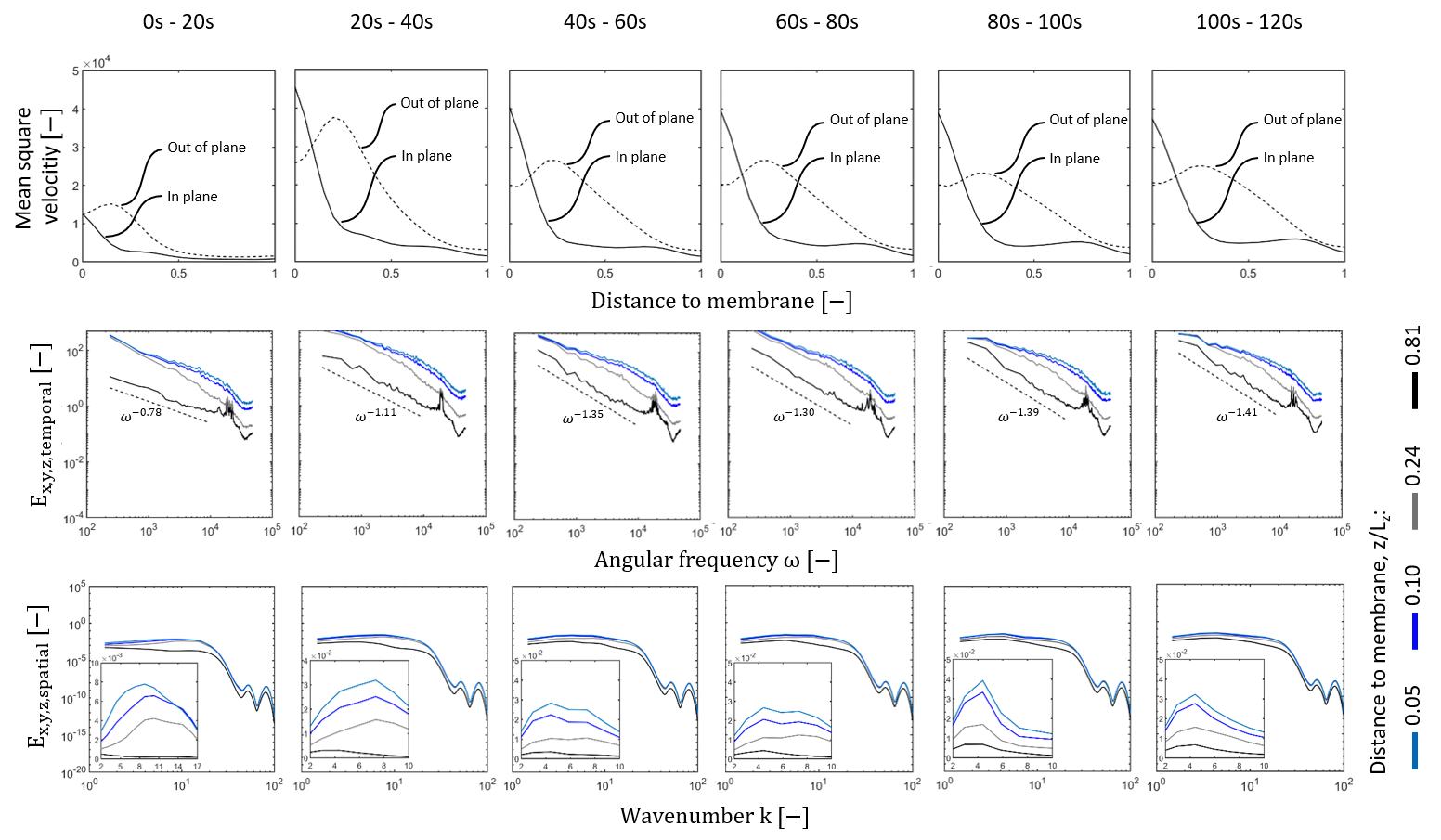}
    \caption{Temporal development of the mean square velocity, temporal and spatial energy spectra for 10.8~$\cdot$~i\textsubscript{lim}. The first row shows the development of the mean square of one in-plane and the out-of-plane velocity components over the plane-to-membrane distance from 3D experiments between \SI{0}{\s} and \SI{120}{\s}. The second and third row show the development of the temporal and spatial energy spectra between \SI{0}{\s} and \SI{120}{\s}, respectively. Note the changing axes for the graphs showing the linear zoom on the maxima of the spatial energy spectra.}
    \label{fig:tempDevVel}
\end{figure*}

Our set up also allows to analyze the development of the EC velocities, vortex structure, and statistics over time. Fig.~\ref{fig:tempDevEC} and~\ref{fig:tempDevVel} show this development for the case of 10.8~$\cdot$~i\textsubscript{lim} in steps of 20 seconds for the first \SI{120}{\s} of the experiment.

The development of the coherent vortex structures is depicted in Fig.~\ref{fig:tempDevEC}. The first coherent structures already build up at \SI{20}{\s} of the experiment appearing as mostly regular vortex rings. The structures grow in diameter over the course of \SI{80}{\s} while loosing their regular size and shape. Up to this stage, the velocity in most centers of coherent structures is directed towards the electrode. The velocity field becomes more chaotic between \SI{100}{\s} and \SI{120}{\s}. In this stage, coherent structures appear in a wide range of shapes and sizes with velocities directed both ways.

Fig.~\ref{fig:tempDevVel} shows the change in one in-plane and the out-of-plane mean square velocities. The distinct velocity profile of EC is not completely build up in the first \SI{20}{\s}. Similar to later stages of EC, the maxima of both velocity components are close to the membrane. However, the out-of-plane velocity is higher than the in-plane component. Both velocity maxima first increase between \SI{20}{\s} to \SI{40}{\s} and decrease to a steady state value thereafter. At this point, the profiles are fully developed.

The above perceptions reflect in the development of the temporal and spatial spectra (Fig.~\ref{fig:tempDevVel}). As explained in the previous section, the local maxima at high frequencies in both the temporal as well as the spatial energy spectra and the following change in slope originate from small-scale noise and the measurement error of PTV for small scales~\cite{Herpin.2008}. Only slight changes are visible in the overall shapes of the spectra for both temporal and spatial analysis. However, some distinct observations can be made. First, the slope of the temporal spectrum  for the plane closest to the electrode ($0.8 \cdot L_z$) changes between \SI{0}{\s} and \SI{40}{\s} from $\omega^{-0.79}$ to $\omega^{-1.35}$ indicating that the energy reaching the region closer to the electrode dissipates faster when EC is fulls developed. Second, the energy levels in both spectra increase from \SI{0}{\s} to \SI{20}{\s} and decrease again after \SI{40}{\s} as already seen in the mean square velocities. Third, the position of the maxima of the 2D spatial spectra change to smaller wavenumbers for all planes over time. The maxima positions shift from 11.6~$k$ to 5.8~$k$ between 0s and 100s as can be seen in the linear zoom of each spectrum. The shift of the maxima correlates to increasing contributions to the energy of vortex structures with larger length scales in the respective planes. The structures with the highest energy share grow from $0.61 \cdot L_z$ to $1.43 \cdot L_z$ over the course of \SI{80}{\s}. The negligible changes in all graphs confirm that the experiment was in a statistical steady state after about \SI{40}{\s}.





\section*{Acknowledgments}
This work received funding from the European Research Council (ERC) under the European Union’s Horizon~2020 research and innovation program (grant agreement no. 694946). The DFG also supported this work through the project SFB 985 Functional Microgels and Microgel Systems in project B6. The measurements were conducted with a high-speed stereomicroscope PIV funded by the Major Research Instrumentation Program according to Art. 91b GG in the Research Building NW1481006 “NGP2 – Center for Next Generation Processes and Products”. M.W. acknowledges DFG funding through the Gottfried Wilhelm Leibniz Award 2019 (grant ID = WE 4678/12-1) and the support from the Alexander-von-Humboldt foundation. The authors acknowledge F. Roghmans, G. Linz, A. Kalde, J. Lölsberg, A. Cinar, K. Baitalow, and D. Wypysek for their input and the fruitful discussions.

\section*{Author contributions}
\textbf{Felix Stockmeier:} Conceptualization, Methodology, Formal analysis, Investigation, Data Curation, Writing - Original Draft, Visualization, Project administration.
\textbf{Michael Schatz:} Formal analysis, Investigation, Visualization.
\textbf{Malte Habermann:} Formal analysis, Investigation, Visualization.
\textbf{John Linkhorst:} Writing - Review \& Editing, Project administration, Supervision.
\textbf{Ali Mani:} Interpretation of the data; Writing - Review \& Editing .
\textbf{Matthias Wessling:} Conceptualization, Resources, Project administration, Supervision, Writing - Review \& Editing, Funding acquisition.

\section*{Competing interests}
The authors declare that there are no competing interests.

\bibliographystyle{elsarticle-num-names}
\bibliography{literature_JMS}





\end{document}